\documentclass[12pt,a4paper,final]{iopart}
\expandafter\let\csname equation*\endcsname\relax
 \expandafter\let\csname endequation*\endcsname\relax
\usepackage{graphicx,amsmath}
\usepackage[breaklinks=true,colorlinks=true,linkcolor=blue,urlcolor=blue,citecolor=blue]{hyperref}
\usepackage{iopams} 
\begin{document}
\title{Characterization of conditional state-engineering quantum processes by coherent state quantum process tomography}
\author{Merlin Cooper, Eirion Slade, Micha{\l} Karpi{\'n}ski, Brian J. Smith}
\address{Clarendon Laboratory, University of Oxford, Parks Road, Oxford, OX1 3PU, UK}
\ead{m.cooper1@physics.ox.ac.uk}
\begin{abstract}
Conditional quantum optical processes enable a wide range of technologies from generation of highly non-classical states to implementation of quantum logic operations. The process fidelity that can be achieved in a realistic implementation depends on a number of system parameters. Here we experimentally examine Fock-state filtration, a canonical example of a broad class of conditional quantum operations acting on a single optical field mode. This operation is based upon interference of the mode to be manipulated with an auxiliary single-photon state at a beam splitter, resulting in the entanglement of the two output modes. A conditional projective measurement onto a single photon at one output mode heralds the success of the process. This operation, which implements a measurement-induced nonlinearity, is capable of suppressing particular photon-number probability amplitudes of an arbitrary quantum state. We employ coherent-state process tomography to determine the precise operation realized in our experiment. To identify the key sources of experimental imperfection, we develop a model of the process and identify three main contributions that significantly hamper its efficacy. The reconstructed tensor is compared with a model of the process taking into account sources of experimental imperfection with fidelity better than $0.95$. This enables us to identify three key challenges to overcome in realizing a filter with high fidelity -- namely the single-photon nature of the auxiliary state, high-mode overlap, and the need for number resolving detection when heralding. The results show that the filter does indeed exhibit a nonlinear response as a function of input photon number and preserves the phase relation between Fock layers of the output state, providing promise for future applications.
\end{abstract}
\maketitle
\section{Introduction}
\label{sec:intro}
Optics is an ideal platform for encoding, manipulating and transmitting quantum information -- photons do not interact strongly with the environment enabling effectively long coherence times that can be harnessed for long-distance communications \cite{O'Brien:NatPhoton:09} and there is extremely low thermal excitation of the optical field at room temperature effectively eliminating background noise. However, the first apparent advantage also poses a serious challenge, namely that photon-photon interactions -- which are crucial for conditional logic operations such as two-qubit gates -- are extremely weak and require currently infeasible optical nonlinear optical interaction strength \cite{D'Ariano:PRA:00,Nielsen::10}. In 2001 Knill, Laflamme and Milburn proposed the concept of linear optics quantum computation (LOQC) \cite{Knill:Nature:01}. In this approach, a large probabilistic nonlinearity could be invoked with the aid of ancilla photons and projective measurements -- thus enabling implementation of photonic logic gates. 

Outside the realm of quantum computation, similar probabilistic interactions can be used to arbitrarily manipulate the modal properties \cite{Resch:PRA:04} or photon-number statistics of light \cite{Resch:PRL:02}. The basic building block of most schemes is the post-selected beam splitter \cite{Dakna:EurPhysJD:98,Hofmann::02}. Such schemes employ a non-classical ancilla state to generate entanglement between the output modes of a beam splitter \cite{Kim:PRA:02}. Performaning a measurement on one output mode of the beam splitter projects the other mode into a state that depends on the specific measurement outcome. This concept has given rise to a host of quantum optical state engineering protocols including photon subtraction \cite{Namekata:NatPhoton:10,Neergaard-Nielsen::11}, photon addition \cite{Dakna:PRA:99,Pegg:PRL:98}, cat-state generation \cite{Outjoumtsev:Nature:07,Ourjoumtsev:Science:06}, quantum scissors \cite{Ozdemir:PRA:01}, photon catalysis \cite{Lvovsky:PRA:02,Bartley:PRA:12}, noiseless amplification \cite{Barbieri:LaserPhysLett:11,Zavatta:NatPhoton:11,Ferreyrol:PRA:11}, entanglement distillation \cite{Takahashi:NatPhoton:10,Kurochkin:PRL:14} and Fock state filtration \cite{Sanaka:PRL:06,Resch:PRL:07}. A thorough review of such photon-level manipulations on travelling modes of light can be found in reference \cite{Jim:NJP:08}.

Experimentally, photon-level manipulations of a single optical mode are impeded by non-ideal ancilla states, poor overlap between ancilla and target modes, and limited availability of low-loss photon-number-resolving detection. In this article a conditional quantum process representative of such measurement-induced nonlinear operations is experimentally investigated. A model of the specific scheme, known as Fock state filtration (FSF) \cite{Sanaka:PRL:06,Resch:PRL:07}, is developed in the presence of experimental imperfections, which is then compared with experimental characterization of a realistic implementation via quantum process tomography. We begin in Section 2, by developing a realistic model of the FSF process that incorporates three central experimental imperfections -- non-number resolving herald detection, mixed state ancilla, and mode mismatch between ancilla and target modes of the FSF. The experimental implementation is then described in Section 3, in which avalanche photodiode detectors and heralded single photons are used to implement the filter. Then in Section 4, the approach to characterize the FSF with coherent-state quantum process tomography is detailed and the results are compared with those predicted by the model developed. In Section 5 we summarize the results and draw conclusions that indicate directions for future research.

\section{Fock state filtration}
\label{sec:model}
\begin{figure}
	\centerline{\includegraphics[width=1.00\textwidth]{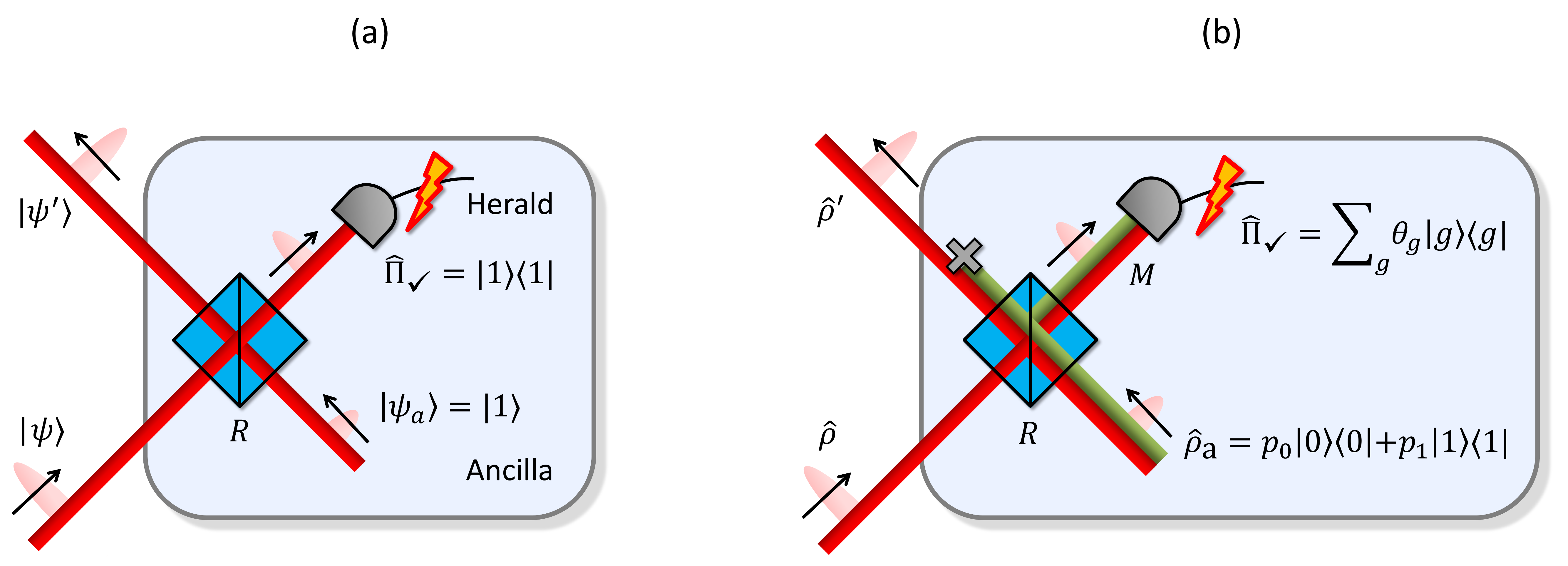}}
	\caption{(a) Schematic of the ideal Fock state filtration quantum operation. Input (pure) quantum state $|\psi\rangle$ is combined with an ancilla single photon $|1\rangle$ at a beam splitter of reflectivity $R$. One output mode is registered by a photon-number-resolving detector, which conditions the operation. Detection of a single photon heralds the success of Fock state filtration, whereby the output state $|\psi'\rangle$ is generated. (b) Schematic of the non-ideal Fock state filtration operation with three sources of imperfection accounting for realistic heralding detection, non-ideal ancilla state preparation and the multimode nature of the heralding detector.} 
	\label{fig:fsf_ideal_model}
\end{figure}
Fock state filtration was proposed by Sanaka \emph{et al.} in 2006 \cite{Sanaka:PRL:06} and is schematically identical to the single-photon catalysis demonstrated by Lvovsky \emph{et al.} in 2002 \cite{Lvovsky:PRL:02}. A simplified depiction of the ideal FSF quantum operation is shown in figure \ref{fig:fsf_ideal_model}(a). An arbitrary pure input quantum state $|\psi\rangle$ is combined with an ancilla single photon $|1\rangle$ at a beam splitter of reflectivity $R$. One output mode of the beam splitter is registered by a photon-number-resolving detector. Only events corresponding to detection of one photon in this output conditioning mode are accepted; whereby the output state $|\psi'\rangle$ is generated in the other mode. Thus detection of a single-photon in the conditioning mode heralds the success of the filter operation. The filtered state $|\psi'\rangle$ contains a `hole' \cite{Baseia:PRA:98} in the photon-number distribution for particular choices of the beam splitter reflectivity $R$ \cite{Sanaka:PRL:06}. 

The term catalysis is sometimes used to describe such an arrangement because internal to the operation, an ancilla single photon is both utilized in the process and subsequently released upon successful implementation of the process heralded by detection of a single photon in the trigger mode of the beam splitter output \cite{Lvovsky:PRL:02,Bartley:CLEO:12,Bartley:PRA:12}. However, owing to quantum interference between the possible paths leading to a single-photon in the conditioning mode, the input state $|\psi\rangle$ is transformed in a non-trivial manner to a different state $|\psi'\rangle$. FSF and the more general catalysis variants \cite{Lvovsky:PRL:02,Bartley:PRA:12} have been shown to be applicable to generate a range of non-classical states of light including photon-number states \cite{Sanaka:PRA:05} and Schr\"{o}dinger cat states \cite{Dakna:PRA:97,Bartley:PRA:12}. Furthermore, it has been proposed that information can be encoded and stored within `holes' in the photon-number distribution \cite{Malbouisson:PhysLettA:01}. Such optical states with one or more `holes' in their photon-number distribution are necessarily non-classical \cite{Baseia:PRA:98}, opening a new platform to study non-classical behaviour. Finally, FSF has been demonstrated as a technique for generating entanglement \cite{Resch:PRL:07} and may be useful as a non-Gaussian operation \cite{Genomi:PRA:10}, a central resource for continuous-variable quantum information processing \cite{Braunstein:RevModPhys:05,Andersen:LaserPhotonRev:10}, when photon subtraction does not suffice.

For the ideal Fock state filter a pure input state $|\psi\rangle$ in the photon-number basis transforms as
\begin{align}
|\psi\rangle=\sum_{n=0}^{\infty}C_n |n\rangle\Rightarrow \mathcal{N}\sum_{n=0}^{\infty}{R}^{(n-1)/2}\left[R-n(1-R)\right]C_n|n\rangle=|\psi'\rangle, \label{eq:fsf_ideal_state}
\end{align}
where $\mathcal{N}$ is a renormalization factor associated with the probabilistic nature of the filter operation. There are two important properties of the state transformation in equation \ref{eq:fsf_ideal_state}. For certain values of the beam splitter reflectivity $R$ the probability amplitude to have $n$ photons in the output state $|\psi'\rangle$ is equal to zero. This occurs for values of $R$ such that $R={n}/({n+1})$. Thus FSF can selectively remove a particular Fock layer $|n\rangle$ from the input state by choosing the beam splitter reflectivity appropriately. Hence FSF can be thought of as a conditional $n$-photon absorber \cite{Sanaka:PRL:06,Resch:PRL:07}, reflecting the measurement induced nonlinearity of this operation. Furthermore, the output state $|\psi'\rangle$ cannot contain any population in Fock layers which were not populated in the original state. Thus the FSF operation cannot re-populate Fock layers which have been filtered. This property enables preparation of Fock states from input coherent or thermal states by implementing a series of FSF operations with different beam splitter reflectivities -- successively filtering out all but one targeted Fock layer \cite{Sanaka:PRA:05}.

Ideal operation of the filter requires a pure ancilla single-photon state perfectly matched to the optical mode to be filtered and conditioning with a perfect photon-number-resolving detector. However, both are currently challenging to requirements to meet in the laboratory. Motivated by this we develop a more realistic model of the FSF operation taking into account three key factors impacting the performance of FSF. Development of such a realistic model allows diagnosis of sources of imperfection in a realistic device exploiting such quantum operations. Figure \ref{fig:fsf_ideal_model}(b) shows the realistic FSF operation including the three sources of imperfection: 1) projection onto one photon in the heralding mode is replaced by a more general positive operator-valued measure (POVM) element describing the conditioning event denoted $\hat{\Pi}_{\checkmark}$, assumed to be diagonal in the photon-number basis; 2) impurity of the ancilla single-photon state in terms of admixture of the vacuum state; 3) the multimode nature of the heralding detector. These are addressed in turn in the proceeding sections, culminating in a realistic model for FSF.

\subsection{Realistic heralding detector}
Ideally the filter only operates when precisely one photon is present in the heralding output mode of the beam splitter, figure \ref{fig:fsf_ideal_model}(b), which corresponds to a successful heralding event being associated with projection onto a single-photon state with a POVM element of the form $\hat{\Pi}_{\checkmark}=|1\rangle\langle1|$. To understand the effect of a general, diagonal POVM element of the form $\hat{\Pi}_{\checkmark}=\sum_g \theta_g |g\rangle\langle g |$ on the filter operation, input-output matrix elements of the FSF beam splitter with the form $\langle n+1-g,g|\hat{U}|n,1\rangle$ are considered. Here $\hat{U}$ is the unitary transformation associated with the beam splitter \cite{Hofmann::02}, $|n,1\rangle$ is the input state to the beam splitter with an $n$-photon state $|n\rangle$ in the target mode and a single-photon state $|1\rangle$ in the ancilla mode, and $| n+1-g,g\rangle$ is the ouptut state of the beam splitter with $g$ photons in the trigger mode and $n+1-g $ in the heralded mode. This matrix element gives the probability amplitude to have $g$ photons in the heralding mode and therefore by the unitarity of the process, which preserves photon number, $n+1-g$ photons in the filter output mode, given $n$ photons are incident in the filter input mode in addition to the ancilla single photon. The probability amplitude for this scenario, denoted $A'_{n+1-g,g}$ is given by
\begin{align}
A'_{n+1-g,g}&=\langle n+1-g,g|\hat{U}|n,1\rangle \nonumber \\
&=\sqrt{\frac{(n-g+1)!g!}{n!}}{R}^{(n-g)/2}(i\sqrt{1-R})^{g-1}\left[ \binom{n}{g-1}R -(1-R)\binom{n}{g} \right], \label{eq:probamp_g}
\end{align}
When the condtioning detector is capable of resolving photon number, projecting the trigger mode onto a state $|g\rangle$ with $g\geq 1$ photons and neglecting dark counts, an arbitrary input superposition state will transform as
\begin{align}
|\psi\rangle=\sum_{n=0}^{\infty}C_n |n\rangle\Rightarrow\mathcal{N}\sum_{n=g-1}^{\infty}C_n A'_{n-g+1,g}|n+1-g\rangle=|\psi'\rangle. \label{eq:trans_g}
\end{align}
However, for a measurement operator of the form $\hat{\Pi}_{\checkmark}=\sum_g \theta_g |g\rangle\langle g |$, a pure input state is transformed to a mixed state. More generally, an impure input state $\hat{\rho}$ transforms as
\begin{align}
\hat{\rho}=\sum_{m,n=0}^{\infty}C_{mn}|m\rangle\langle n|\Rightarrow 
&\sum_{g=1}^{\infty}\theta_g \sum_{m,n=g-1}^{\infty} \nonumber \\ & C_{mn}A'_{m-g+1,g}A'^*_{n-g+1,g}|m+1-g\rangle\langle n+1-g|=\hat{\rho}', \label{eq:6:rho_trans_det}
\end{align}
which is nothing but a mixture of transformations of the form given by equation \ref{eq:trans_g}, weighted by the coefficients $\theta_g$ describing the heralding POVM element in the Fock basis. 

In general a quantum process can be uniquely described by a rank-4 tensor, which relates the matrix elements in the Fock basis of the input and output states through the relation
\begin{align}
[\rho_{\text{out}}]_{jk} = \sum_{m,n} \mathcal{E}_{jk}^{mn} \rho_{mn} \label{eq:x_outputdensity}
\end{align}
where the input density operator $\hat{\rho}$ in the Fock basis, represented by the matrix $\rho_{mn}$, is mapped onto an output density operators $[\rho_{\text{out}}]_{jk}$ through the action of the quantum process.

The process tensor describing the state transformation in equation \ref{eq:6:rho_trans_det} can be derived by considering the action of the process on input coherent states \cite{Rahimi-Keshari:NJP:11}. The process tensor for FSF when employing a general non-number-resolving phase-insensitive heralding detector POVM, denoted $(\mathcal{E}_1)_{jk}^{mn}$, is given by
\begin{align}
(\mathcal{E}_1)_{jk}^{mn}=\sqrt{\frac{j!k!}{m!n!}}&\sum_{g=1}^{\infty}\theta_g g! {R}^{(j+k-2)/2} ({1-R})^{(g-1)} \nonumber \\ &\times 
\left[\binom{j+g-1}{g-1}R-(1-R)\binom{j+g+1}{g} \right] \nonumber \\ &\times \left[ \binom{k+g-1}{g-1}R - (1-R)\binom{k+g-1}{g} \right] \nonumber \\ &\times \delta_{m,j+g-1}\delta_{n,k+g-1}. \label{eq:6:fsf_nonideal_povm_tensor}
\end{align}

\subsection{Ancilla state heralding efficiency}
In the realistic FSF operation, depicted in figure \ref{fig:fsf_ideal_model}(b), the ancilla state, $\hat{\rho}_{\text{a}}$,  is modelled as a mixture of the single-photon and vacuum states such that
\begin{align}
\hat{\rho}_{\text{a}}=(1-\eta_{\text{H}})|0\rangle\langle 0|+ \eta_{\text{H}}|1\rangle\langle 1 |, \label{eq:admixed_ancilla}
\end{align}
where $\eta_{\text{H}}$ is the ancilla single-photon state heralding efficiency -- a commonly used parameter that describes the efficiency with which a single photon heralded from, for example, a spontaneous parametric down-conversion (SPDC) source, can be matched to a desired optical mode \cite{Cooper:OpEx:13}. To model the effect of non-unit heralding efficiency we take account of the heralding events which occur when the ancilla state is in the vacuum state. The probability amplitudes are derived for input Fock states and are given by
\begin{align}
A''_{n-g,g}&=\langle n-g,g|\hat{U}|n,0\rangle=\sqrt{\binom{n}{g}}{R}^{(n-g)/2}(i\sqrt{1-R})^g \label{eq:probamp_g_0}.
\end{align}
This is the amplitude to find $g$ photons at the trigger output mode of the beam splitter and $n-g$ photons in the other output when the target input mode has $n$ photons and the ancilla input mode occupies the vacuum state. In exact analogy to equation \ref{eq:6:rho_trans_det}, for a realistic heralding detector described by POVM element $\hat{\Pi}_{\checkmark}=\sum_g \theta_g |g\rangle\langle g |$ and vacuum ancilla state, an arbitrary input state described by density matrix $\hat{\rho}$ will transform as
\begin{align}
\hat{\rho}=\sum_{m=0}^{\infty}\sum_{n=0}^{\infty}C_{mn}|m\rangle\langle n|\Rightarrow 
\sum_{g=1}^{\infty}\theta_g \sum_{m=g}^{\infty}\sum_{n=g}^{\infty} C_{mn}A''_{m-g,g}A''^*_{n-g,g}|m-g\rangle\langle n-g|=\hat{\rho}', \label{eq:6:rho_trans_her}
\end{align}
where dark counts in the heralding detector are neglected. The process tensor describing the state transformation of equation \ref{eq:6:rho_trans_her} is can be obtained in the same manner as for equation \ref{eq:6:fsf_nonideal_povm_tensor}. Denoted $(\mathcal{E}_0)_{jk}^{mn}$, the tensor elements associated with the vacuum ancilla and non-number resolving phase-insensitive conditioning detector is given by
\begin{align}
(\mathcal{E}_0)_{jk}^{mn}=\frac{1}{\sqrt{m!n!}}\sum_{g=1}^{\infty}\theta_g\sqrt{(j+g!)(k+g)!}&\sqrt{R}^{j+k}\sqrt{1-R}^{2g}\sqrt{\binom{j+g}{g}\binom{k+g}{g}} \nonumber \\ 
&\times\delta_{m,j+g}\delta_{n,k+g}.
\end{align}
By the linearity of quantum mechanics, a mixed ancilla state $\hat{\rho}_{\text{a}}$ will lead directly to a mixture of the two process tensors derived in this section and the preceding section, i.e. a mixture of $\mathcal{E}_0$ and $\mathcal{E}_1$, weighted according to the heralding efficiency $\eta_{\text{H}}$. Thus the process map for realistic FSF is given by
\begin{align}
\mathcal{E}_{\text{FSF}}=\eta_{\text{H}}\mathcal{E}_1+(1-\eta_{\text{H}})\mathcal{E}_0. \label{eq:6:proc_ten_det_her}
\end{align}

\subsection{Multimode heralding detector}
Photon-counting detectors such as avalanche photodiodes (APDs) and time-multiplexed detectors (TMDs) \cite{Achilles:OptLett:03} are general spatially, temporally and polarization multimode. A well-defined spatial and polarization mode are typically selected by using a single-mode fiber and polarizing element in front of the detector to filter a single spatial and polarization mode. However, since the response time of a typical photon-counting detector is often far greater than the optical pulse duration, such detectors are inherently temporally multimode \cite{Silberhorn:ContPhys:07}. Tualle-Brouri \emph{et al.} studied the impact of using a multimode heralding detector in a scheme to prepare Schr\"{o}dinger cat states by photon subtraction from squeezed states of light \cite{Tualle-Brouri:PRA:09}. It was shown that for such a scheme, the problem could be reduced to two effective modes, thus aiding analysis of the multimode effects. In the realistic FSF scheme depicted in figure \ref{fig:fsf_ideal_model}(b) a similar two-mode decomposition may be performed. It is assumed that the FSF input state $\hat{\rho}$ (e.g. a coherent state) occupies a single well-defined spatial-temporal-polarization mode. The impact of the multimode heralding detector arises due to the potentially multimode nature of the ancilla state $\hat{\rho}_{\text{a}}$, which could contribute a false heralding event. 

Assuming the FSF input state $\hat{\rho}$ occupies a single mode 
the problem can be decomposed into two effective modes: 1) the mode defined by the input state and 2) the mode orthogonal to this in which the ancilla has some population \cite{Tualle-Brouri:PRA:09}. Ideally, the ancilla state mode would perfectly match the FSF input, in which case the problem reduces to a single mode. In practice, a fraction of the ancilla mode may not be overlapped with the FSF input but may still lead to detection events in the heralding detector. In figure \ref{fig:fsf_ideal_model}(b) this is represented pictorially by the green-shaded part of the ancilla-state mode. A `multimode' parameter $M$ is defined such that
\begin{align}
M=\frac{\eta_{\text{H}}}{\eta'_{\text{H}}},
\end{align}
where $\eta_{\text{H}}$ is the heralding efficiency of the ancilla single-photon state into the same mode as the FSF input and $\eta'_{\text{H}}$ is the heralding efficiency registered by the multimode FSF heralding detector, such that $\eta'_{\text{H}}\geq\eta_{\text{H}}$.
For $M=1$ the problem is single-mode and thus the process tensor of equation \ref{eq:6:proc_ten_det_her} applies. Events where the FSF heralding detector `clicks' due to detection of the ancilla single-photon in the mode which is not overlapped with the input state corresponds to there having been no interaction between the input state $\hat{\rho}$ and the ancilla. In these events the beam splitter will simply attenuate the input state. This can be modelled as a probabilistic attenuation process, where the input state intensity is attenuated by a factor of $1-R$. The process tensor describing attenuation is given by \cite{Rahimi-Keshari:NJP:11}
\begin{align}
(\mathcal{E}_{\text{att}})_{jk}^{mn}=\sqrt{\frac{m!n!}{j!k!}}\frac{{\eta}^{(j+k)/2}(1-\eta)^{m-j}}{(m-j)!}\delta_{m-j,n-k}, \label{eq:6:atten_Tensot}
\end{align}
where $\eta$ is the fraction of the incoming intensity which is passed to the output mode of the beam splitter, i.e. $\eta=R$, figure \ref{fig:fsf_ideal_model}(b).

Thus, the complete process tensor describing the realistic FSF operation as depicted in figure \ref{fig:fsf_ideal_model}(b) is given by
\begin{align}
\mathcal{E}_{\text{FSF}}=M\left( \eta_{\text{H}}\mathcal{E}_1+[1-\eta_{\text{H}}]\mathcal{E}_0 \right)+(1-M)\eta_{\text{det}}R\mathcal{E}_{\text{att}}, \label{eq:model_tensor}
\end{align}
where $R$ is the FSF beam splitter reflectivity and $\eta_{\text{det}}$ is the quantum efficiency of the heralding detector -- both of which must be included explicitly as multiplicative factors for the attenuation component to correctly account for the success probability of this event.

\section{Experimental setup}
\label{sec:setup}
Fock state filtration was first demonstrated experimentally by Sanaka \emph{et al.} in 2006 \cite{Sanaka:PRL:06} and later by Resch \emph{et al.} in 2007 \cite{Resch:PRL:07}. Both experimental demonstrations were similar in that the nonlinear absorption property of the filter was inferred from two- and four-fold coincidence measurements performed at the output of the FSF for incident one- and two-photon input states to the filter. The beam splitter reflectivity $R$ was varied and the Hong-Ou-Mandel visibility \cite{Hong:PRL:87} recorded for the two different input Fock states. Operation of the filter consistent with theory was inferred by noting that for an incident one-photon Fock state the best visibility occurred for $R=1/2$ and for an incident two-photon Fock state for $R=2/3$ \cite{Sanaka:PRL:06,Resch:PRL:07}. Although these experiments go some way towards demonstrating the filter operation, such coincidence measurements are inherently post-selected on preservation of photon-number, and are thus loss insensitive. This masks the effect of non-unit heralding efficiency and to some extent enables a degree of number-resolution in the APD heralding detector. Furthermore, probing the filter with Fock states does not enable preservation of coherence between Fock layers to be verified, which requires input states with a superposition of photon number states, e.g. a coherent state.

The technique of coherent-state quantum process tomography (csQPT) \cite{Lobino:Science:08,Rahimi-Keshari:NJP:11,Anis:NJP:12} does not require the use of difficult to prepare non-classical number-state superpositions to probe a quantum process, but rather readily available coherent states emitted from a stable laser system. Ideal filter operation requires a perfect number-resolving heralding detector and a pure ancilla photon state, perfectly mode-matched to the mode of the quantum state to be filtered. We probe an implementation of the process with realistic components to assess the effect of such imperfections. Thus to verify the model of FSF developed above, equation \ref{eq:model_tensor} , csQPT is performed on an experimental implementation of the FSF operation, enabling reconstruction of an estimate of the process tensor $\mathcal{E}_{\text{FSF}}$ from experimental data.

\begin{figure}
\centerline{\includegraphics[width=1.00\textwidth]{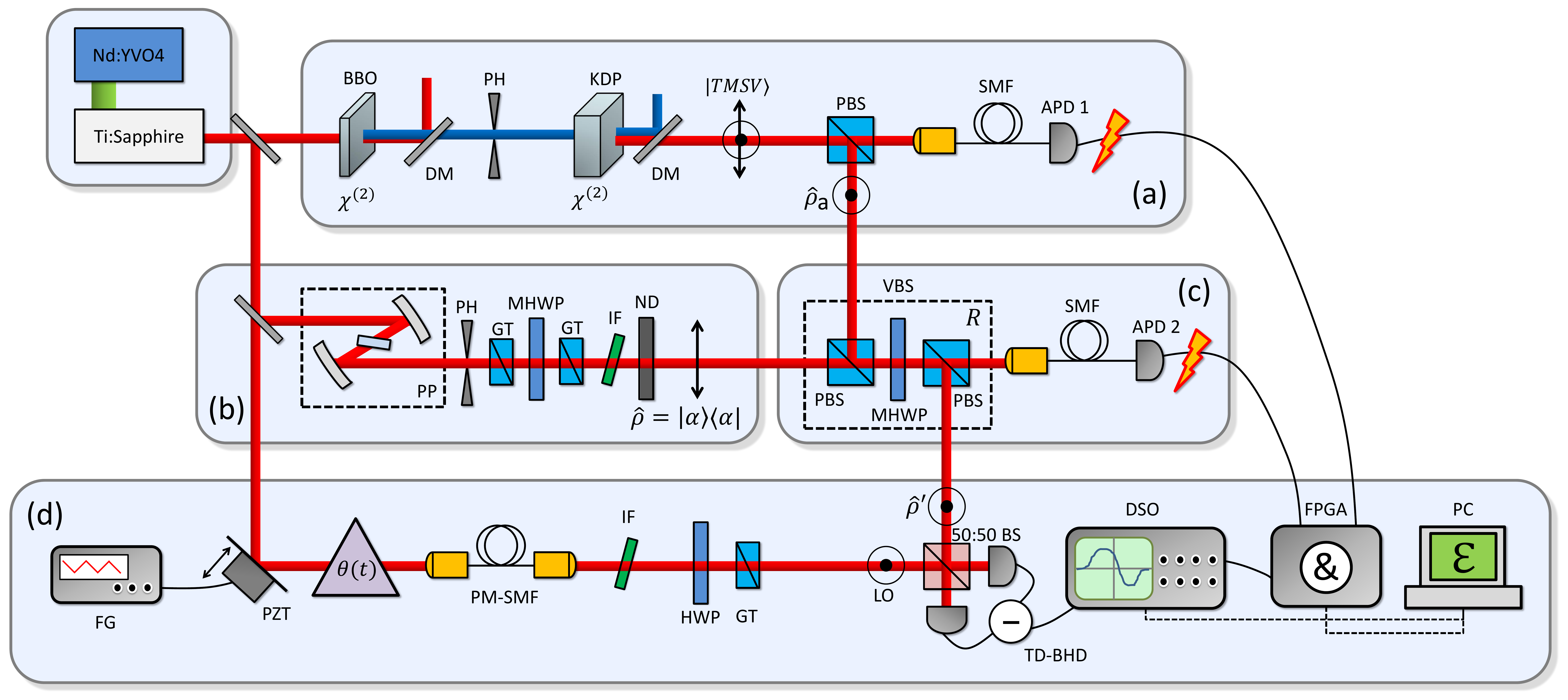}}
\caption{Experimental setup for performing csQPT of the Fock state filter. The output of a Ti:Sapphire oscillator is split into 3 spatial paths to serve as the pump for second-harmonic generation (SHG), the coherent state probes and the local oscillator (LO). (a) The SHG output pumps a potassium dihydrogen phosphate (KDP) crystal where Type-II degenerate spontaneous parametric down-conversion (SPDC) occurs. The output modes from the SPDC are separated at a polarizing beam splitter (PBS). One mode is coupled into a single-mode fiber (SMF) and detected with an avalanche photodiode detector (APD 1) to herald the single-photon ancilla state $\hat{\rho}_{\text{a}}$. {(b)} The repetition rate of the pulse train of coherent state probes is reduced using a pulse picker (PP). Spatial and spectral filtering is achieved using a pinhole (PH) and interference filter (IF) respectively. The probe state amplitude control consists of a half-wave plate (HWP) situated between two Glan-Taylor polarizers (GT) followed by neutral density (ND) filters. {(c)} The coherent state $\hat{\rho}=|\alpha\rangle\langle\alpha|$ and heralded ancilla state $\hat{\rho}_{\text{a}}$ are combined at a variable beam splitter (VBS) constructed from a HWP situated between two PBS. One output mode of the VBS is coupled to a SMF and detected with APD 2 to herald the success of the FSF process. {(d)} The FSF output state $\hat{\rho}'$ is combined with the LO on a 50:50 beam splitter to perform balanced homodyne detection. A digital storage oscilloscope (DSO) triggered from the APDs records the time-domain balanced homodyne detector (TD-BHD) output. The success probability of the FSF operation is recorded with an FPGA. The LO phase $\theta_{\text{LO}}(t)$ is swept by a piezo-electric transducer (PZT), driven with a triangular wave from a function generator (FG). Further symbols and abbreviations are defined in the main text.}
\label{fig:schematic}
\end{figure}
The complete optical schematic of the setup used to perform csQPT of the FSF operation is shown in figure \ref{fig:schematic} and is split into four distinct stages: (a) preparation of the ancilla single photon $\hat{\rho}_{\text{a}}$, (b) preparation of coherent states $\hat{\rho}=|\alpha\rangle\langle\alpha|$ used to probe the FSF operation, (c) the FSF beam splitter and heralding detection and (d) homodyne detection of the FSF output state $\hat{\rho}'$ to perform csQPT. Each stage is detailed in the following sections.

\subsection{Ancilla state preparation}
\begin{figure}
\centerline{\includegraphics[width=1.0\textwidth]{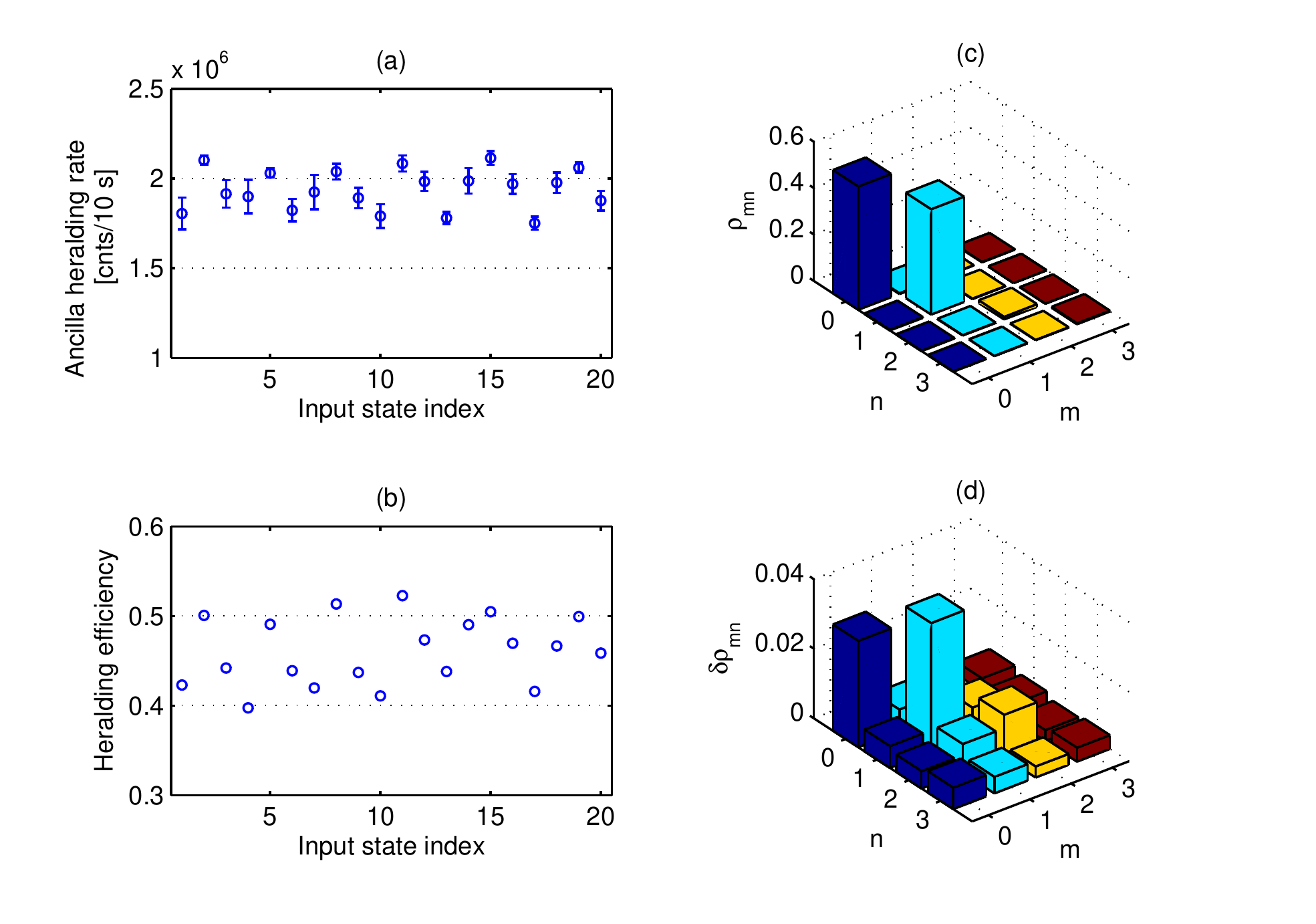}}
\caption{Characterization of the ancilla single-photon Fock state $\hat{\rho}_{\text{a}}$ by homodyne tomography. (a) Heralding rate registered by APD 1, recorded by the FPGA, monitored throughout the FSF QPT experiment for each input probe state. Fluctuations are mainly due to instability of the 415 nm pump beam pointing. (b) Heralding efficiency $\eta_{\text{H}}$ of the ancilla single photon in the spatial-temporal mode examined by the BHD. Fluctuations in the heralding efficiency are correlated with fluctuations in the heralding rate, thus indicating the fluctuations are due to the SPDC pump pointing instability. (c) Reconstructed density matrix and (d) associated errors obtained by performing quantum state tomography of the ancilla state. Higher-order photon-number terms are negligible. The heralding efficiency of the ancilla single-photon state is determined to be $\eta_{\text{H}} =0.45 \pm 0.04$.}
\label{fig:ac}
\end{figure}
The ancilla single-photon state $\hat{\rho}_{\text{a}}$ is derived in a heralded fashion from a spontaneous parametric down-conversion (SPDC) source \cite{Cooper:OpEx:13,Mosley:PRL:08}. An 80~MHz train of 100~fs pulses at 830~nm central wavelength from a Ti:Sapphire oscillator (Spectra-Physics Tsunami) is frequency doubled with a conversion efficiency of approximately 30\% in a 700~$\mu$m long $\beta$-barium borate (BBO) crystal cut for type-I collinear phase matching resulting in 85~fs pulses at 415~nm. The residual fundamental is filtered out using dichroic mirrors and Schott glass filters, resulting in a maximum second-harmonic power of 500~mW. The second-harmonic beam is subsequently spatially filtered with a pinhole and focused into an 8 mm thick potassium dihydrogen phosphate (KDP) crystal cut for Type-II collinear degenerate down conversion \cite{Mosley:PRL:08,Cooper:OpEx:13}. The source produces a nearly two-mode squeezed vacuum state (TMSV), which is described in the photon-number basis as
\begin{align}
|\psi\rangle\approx\sqrt{1-\lambda^2}\left(|0,0\rangle + \lambda |1,1\rangle + \lambda^2 |2,2\rangle + O(\lambda^3)\right), \label{eq:TMSV_approx}
\end{align}
where the parameter $\lambda$ is related to the squeezing parameter and depends on the pulse energy of pump and $|m,n\rangle$ describes the state with $m$ ($n$) photons in the trigger (ancilla) mode. Since the KDP crystal is cut for Type-II phase matching, the trigger and ancilla modes are orthogonally polarized and easily separated using a polarizing beam splitter (PBS). The trigger mode is coupled into a single-mode fiber (SMF) to select a well-defined spatial mode from the SPDC source and detected with an avalanche photodiode detector (APD) (Perkin-Elmer SPCM-AQ4C). Although the APD is a binary detector, since $\lambda$ is extremely small in the experiment ($\lambda\approx0.07$), when the APD registers a click the ancilla mode collapses to a state very closely approximating a single-photon Fock state with negligible higher-order photon number terms. 

In practice, it is generally difficult to exactly mode match the heralded ancilla single-photon state both spatially and temporally with, for example, the local oscillator (LO) used to perform balanced homodyne detection \cite{Cooper:OpEx:13}. Even if the SPDC source produces only a single two-mode squeezed state in the spatial modes examined (i.e. the joint spectral amplitude for the trigger and ancilla fields is factorable \cite{Mosley:PRL:08}) there is no guarantee the ancilla mode can be well matched to the desired mode. This leads to an ancilla state overlapped with the target mode consisting of an admixture of the heralded single-photon state with vacuum, as given in equation \ref{eq:admixed_ancilla},
where $\eta_{\text{H}}$ is the heralding efficiency into the target mode. During the experiment $\eta_{\text{H}}$ is monitored by setting the HWP in the FSF beam splitter to direct the ancilla state directly to the balanced homodyne detector. This enables quantum state tomography (QST) of the ancilla state from which $\eta_{\text{H}}$ can be extracted, figure \ref{fig:ac}. The heralding efficiency during the experiment was found to be $\eta_{\text{H}}=0.45\pm0.04$.

To determine the parameter $M$ characterizing the effect of using a multimode heralding detector to condition the FSF operation, the heralding efficiency $\eta_{\text{H}}'$ in the FSF heralding mode is determined using APD 2, figure \ref{fig:schematic}. The HWP in the FSF beam splitter is set to direct the ancilla state to conditioning detector APD 2 where the heralding efficiency $\eta_{\text{H}}'$ is directly monitored using a coincidence counting program implemented in a field-programmable gate array (FPGA). After correcting for the APD efficiency $\eta_{\text{APD}}=0.45$ the heralding efficiency registered by APD 2 is determined to be $\eta_{\text{H}}'=0.62$. This gives a value of the multimode parameter $M={0.45}/{0.62}=0.73$.

\subsection{Probe state preparation}
A small fraction of the original 830~nm oscillator output is split off to serve as both the local oscillator (LO) for performing balanced homodyne detection and to prepare the coherent state probes to perform csQPT. The probe state repetition rate is reduced by a factor of 20 using a pulse picker (PP) (A·P·E Angewandte Physik \& Elektronik GmbH pulseSelect kit) and subsequently spatially filtered with a pinhole (PH) and spectrally filtered with an interference filter (IF) (Semrock LL01-830-12.5) to match the mode of the ancilla state $\hat{\rho}_{\text{a}}$. The probe state amplitude control consists of a motorized half-wave plate (MHWP) situatated between two Galn-Taylor (GT) polarizers, figure \ref{fig:schematic}(b). The repetition rate reduction of the probe states serves two purposes. Firstly, it prevents the conditioning detector (APD 2), which heralds the FSF filter operation, from saturating when relatively bright coherent states are used to probe the filter. Secondly, the reduced probe state repetition rate enables access to a set of `dark' pulses in the vacuum state with which to contemporaneously calibrate the BHD during data acquisition. This ensures the acquired quadrature data are free from systematic errors due to drifting of the BHD \cite{Cooper:JMO:13}. The input probe state amplitude $|\alpha|$ is accurately measured throughout the experiment by performing QST of the probe states that do not undergo the FSF operation.

\subsection{FSF beam splitter and balanced homodyne detection}
The variable beam splitter (VBS) for performing FSF consists of a motorized (MHWP) and two polarizing beam splitters (PBSs). The ancilla state $\hat{\rho}_{\text{a}}$ is combined into the same spatial mode as the coherent probe state $|\alpha\rangle$ with the first PBS. The HWP followed by the second PBS enables any reflectivity $R$ to be selected. One output mode of the second PBS is coupled to an APD using a SMF. This serves as the conditioning mode for the FSF. The conditioning measurement is described by the POVM element for the APD `click' event given by $\hat{\Pi}_{\text{click}}=\sum_{g=0}^{\infty}\theta_g|g\rangle\langle g|$ where $\theta_g=1-(1-\eta_{\text{apd}})^g$ and $\eta_{\text{APD}}=0.45$ is the APD efficiency \cite{Silberhorn:ContPhys:07}.

The FSF output state $\hat{\rho}'$ is detected with a BHD by interference with the LO on 50:50 beam splitter followed by a pair of photodiodes with photocurrents directly subtracted and amplified \cite{Cooper:JMO:13}. The LO spatial mode is defined by a short (15~cm) polarization-maintaining single-mode fiber (PM-SMF), and the spectral mode is defined with an interference filter (IF) (Semrock LL01-830-12.5), such that the LO mode is well-matched with the FSF output state mode. The LO phase $\theta$ is swept by modulating one of the interferometer mirrors using a piezo-electric transducer (PZT). The BHD output voltage $V_{\text{BHD}}$ is proportional to a sample of the generalized quadrature of the FSF output state, $V_{\text{BHD}}\propto\hat{X}_{\theta}=\frac{1}{\sqrt{2}}\left(\hat{a}e^{i\theta}+\hat{a}^{\dagger}e^{-i\theta}\right)$. $V_{\text{BHD}}$ is recorded by a digital storage oscilloscope (DSO). Data acquisition from the BHD is triggered by three events occurring in coincidence: 1) a control pulse indicating the pulse picker has generated a probe state $|\alpha\rangle$, 2) an ancilla photon heralding event from APD 1 and 3) a FSF success heralding event from APD 2. The internal pattern trigger function of the oscilloscope is used to automatically select these events. For each probe state $|\alpha\rangle$ 40 sets of 8000 frames are recorded. Each frame consists of one pulse containing a single sample of the output state $\hat{\rho}'$ and 9 pulses containing single samples of the vacuum state to calibrate the BHD \cite{Cooper:JMO:13}. Thus a total of $3.2\times10^5$ samples are recorded for each probe state. In total, quadrature samples for 20 probe states with amplitude $|\alpha|$ ranging from 0.1 to 1.5 in approximately equal steps were recorded. The LO phase $\theta$ is extracted from the output state average quadrature value over fixed time intervals \cite{Bimbard:NatPhoton:10}.

\section{Reconstruction of FSF process tensor}
\begin{figure}
\centerline{\includegraphics[width=1.0\textwidth]{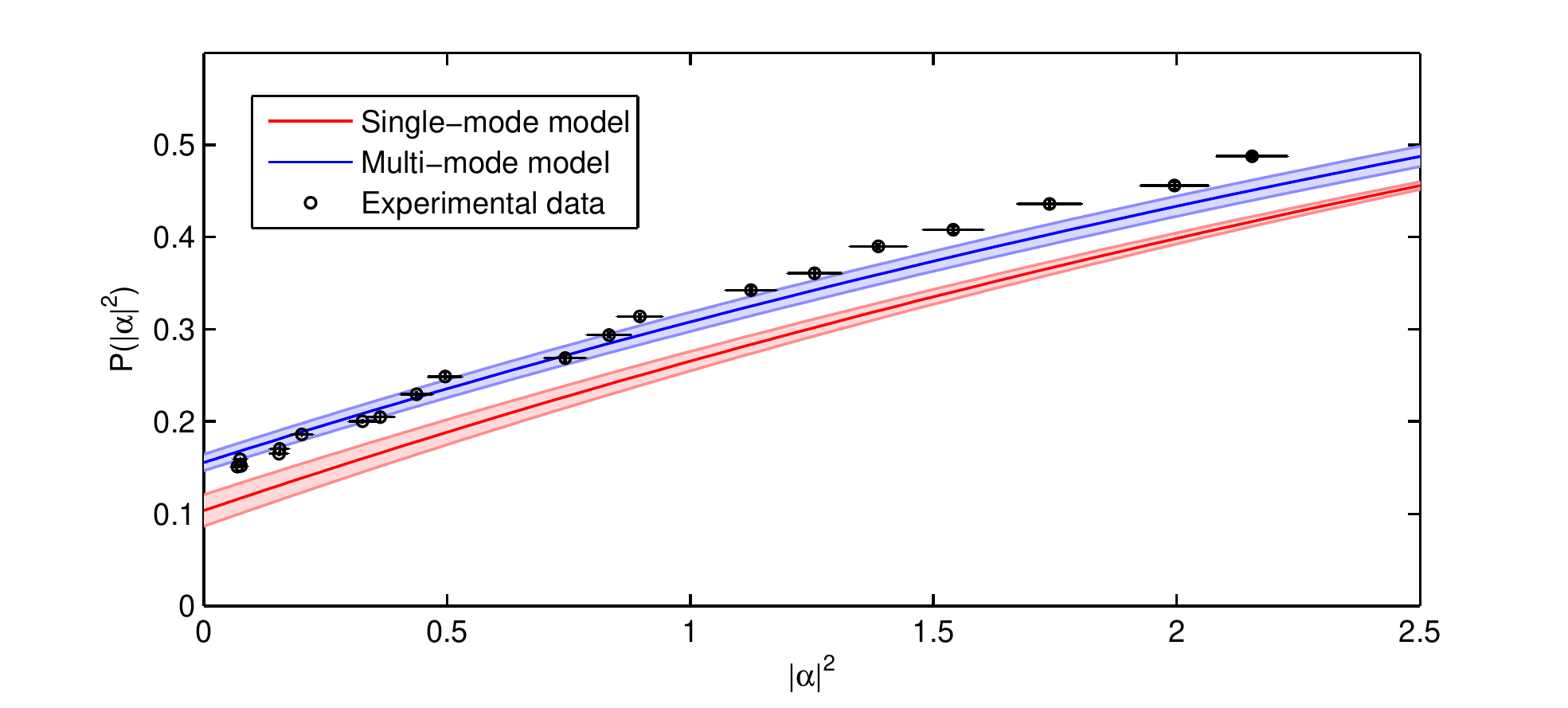}}
\caption{Success probability of the FSF operation as a function of the input coherent state intensity $|\alpha|^2$. Experimental values recorded by the FPGA (black circles), prediction of multimode model of FSF (blue) and single-mode model (red), developed in section \ref{sec:model}.}
\label{fig:success_probability}
\end{figure}
Since FSF is a conditional process, heralded by the simultaneous generation of an ancilla single-photon state $\hat{\rho}_{\text{a}}$ and the detection of a photon in the FSF heralding mode, it is necessary to record the success probability of the operation, $P(\alpha)$, which is in general a function of the input probe state, to perform csQPT \cite{Anis:NJP:12,Paris::04}. The data captured by the BHD do not contain information about the success probability. The success probability is monitored using a coincidence counting program implemented in a FPGA, figure \ref{fig:schematic}(d). The output TTL pulses from APD 1 and APD 2 are split between the DSO for triggering data acquisition from the BHD and the FPGA to record the success probability. The measured success probability as a function of the probe state intensity $|\alpha|^2$ is shown in figure \ref{fig:success_probability} in addition to the predicted success probability derived from the single-mode and multimode models of FSF presented in section \ref{sec:model}, where the parameter values measured in section \ref{sec:setup} are used. The experimentally measured success probability $P(|\alpha|^2)$ is consistent with the multimode model of FSF -- whilst the single-mode model predicts a consistently lower probability of success. The emergent discrepancy between the multimode model success probability and the experimental success probability for $|\alpha|^2>1.25$ is an indication that the probe state mode and LO mode are not perfectly overlapped -- as was assumed in order to decompose the problem into two effective modes. However, in principle these modes can be made to match perfectly since they are derived from the same laser.

\begin{figure}
\centerline{\includegraphics[width=1.0\textwidth]{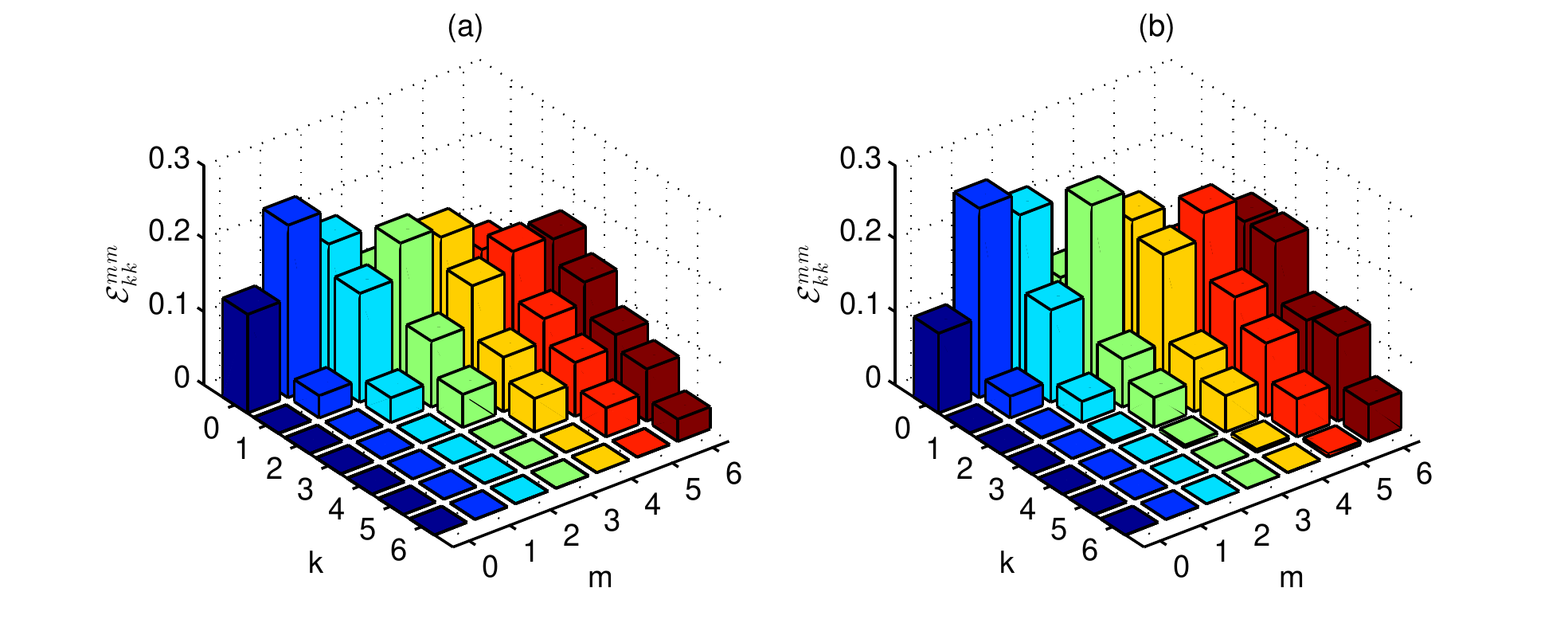}}
\caption{Diagonal elements $\mathcal{E}_{kk}^{nn}$ of the model (a) and reconstructed (b) process tensors for Fock state filtration. The model tensor was calculated according to equation \ref{eq:model_tensor} with parameter values as determined in section \ref{sec:setup}.}
\label{fig:fsf_recon_diagonal}
\end{figure}
The raw quadrature samples for the FSF output states recorded by the BHD are binned with 30 phase bins in the interval $\theta \in [0,\pi]$ and 601 quadrature bins in the interval $X \in [-5,5]$. The reconstruction is performed up to a maximum photon number $n=6$. Thus the reconstructed process tensor is able to predict the evolution of an arbitrary input state $\hat{\rho}=\sum_{m,n=0}^6 \rho_{mn}|m\rangle\langle n|$ on the truncated input Hilbert space $\mathcal{H}$. The maximum-likelihood reconstruction algorithm \cite{Anis:NJP:12}, implemented in MATLAB, took approximately 6.5 hours to perform 150 iterations on a multi-core desktop computer. The iterations are stopped when the change in the likelihood approaches the machine precision corresponding to negligible change of the process tensor elements. Furthermore, dilution of $\mu=0.5$ was used \cite{Rehacek:PRA:2007} to curb oscillations in the likelihood at the start of the reconstruction \cite{Anis:NJP:12}. Figure \ref{fig:fsf_recon_diagonal}(a) shows diagonal elements $\mathcal{E}^{nn}_{kk}$ of the model process tensor (equation \ref{eq:model_tensor}), where the model parameters are those determined in section \ref{sec:setup}, and figure \ref{fig:fsf_recon_diagonal}(b) of the reconstructed process tensor. 

On inspection the model and reconstructed diagonal elements are similar, both in terms of the structure within each input Fock layer and the sum of each input Fock layer, which represents the success probability for a particular input Fock state. A more quantitative analysis is afforded by calculating the fidelity between the reconstructed and model process tensors according to
\begin{align}
F=\text{Tr}\left[\left((\hat{E}_{\text{recon}})^{1/2}\hat{E}_{\text{model}}(\hat{E}_{\text{recon}})^{1/2}\right)^{1/2}\right]^2, \label{eq:6:fidelity_jcoi}
\end{align}
where the processes are represented as operators on the combined input-output Hilbert space $\mathcal{H}\otimes\mathcal{K}$ according to the Jamio{\l}kowski-Choi isomorphism \cite{Paris::04}. So that the fidelity $F$ is bounded such that $0\leq F \leq 1$ it is necessary to normalize each operator $\hat{E}$ \cite{Bongioanni:PRA:10}. This preserves the relative success probability for each input Fock layer and effectively amounts to changing the overall `gain' of the process. The fidelity thus defined is a function of the whole process tensor, i.e. not just the diagonal elements displayed in figure \ref{fig:fsf_recon_diagonal}.

\begin{figure}
\centerline{\includegraphics[width=1.0\textwidth]{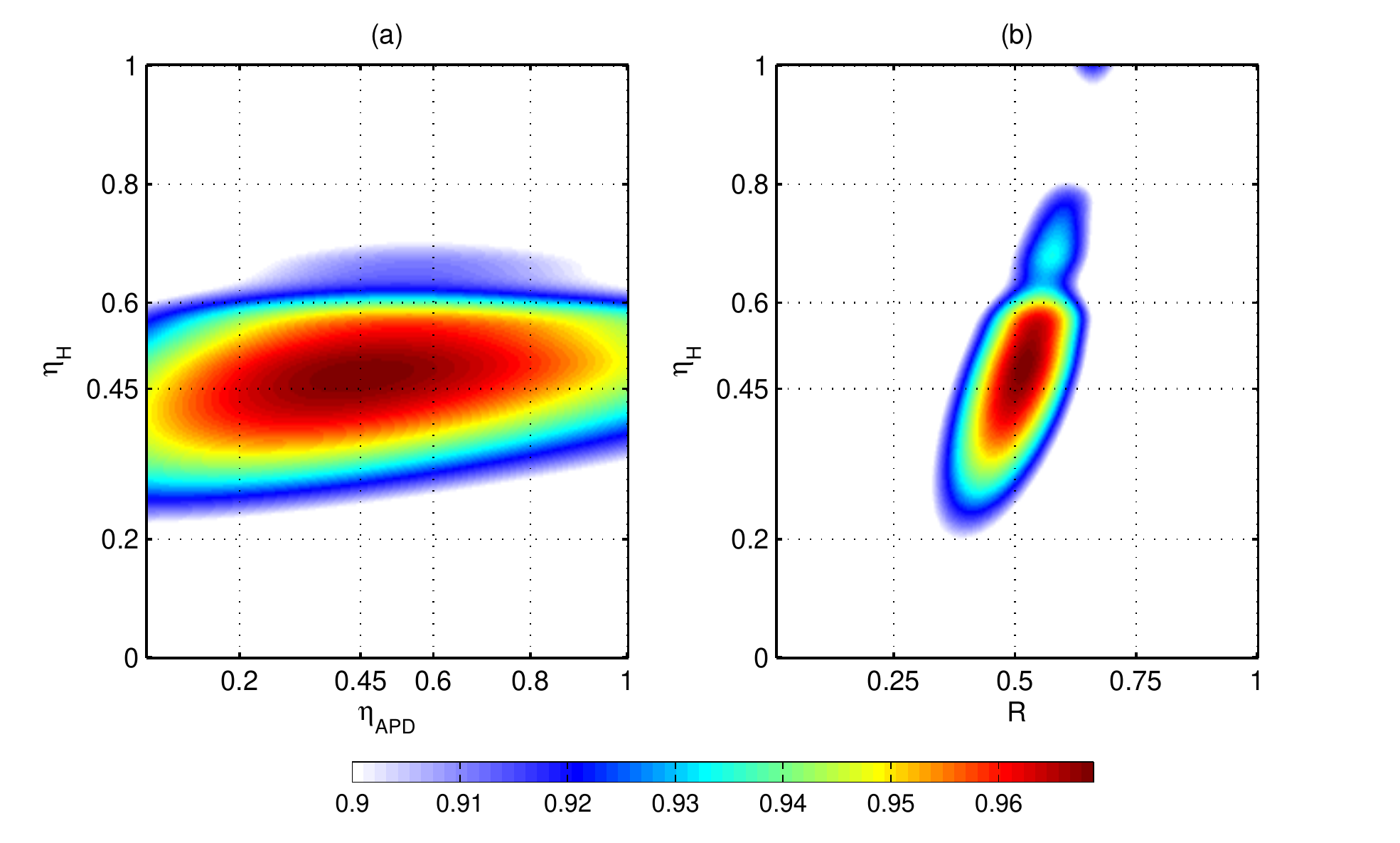}}
\caption{Fidelity between reconstructed tensor and model tensor of FSF as a function of model parameter values. (a) Fidelity as a function of the APD efficiency $\eta_{\text{APD}}$ and heralding efficiency $\eta_{\text{H}}$ for fixed $M=0.73$, $R=0.5$. (b) Fidelity as a function of the beam splitter reflectivity $R$ and heralding efficiency $\eta_{\text{H}}$ for fixed $M=0.73$ and $\eta_{\text{APD}}=0.45$. The fidelity peaks in the region of the parameters values determined in section \ref{sec:setup}. Thus the model tensor provides the best match to the reconstructed tensor for the expected parameter values -- further indicating validity of the model tensor, equation \ref{eq:model_tensor}.}
\label{fig:imagesc}
\end{figure}
The fidelity is a multi-dimensional function of the model parameters and would ideally be maximal for the parameter values experimentally determined in section \ref{sec:setup}. To this end, two cuts of the fidelity as a function of the model parameters were calculated. Figure \ref{fig:imagesc}(a) shows the fidelity as a function of the ancilla state heralding efficiency $\eta_{\text{H}}$ and APD efficiency $\eta_{\text{APD}}$ used in the model of equation \ref{eq:model_tensor}, with the beam splitter reflectivity $R=0.5$ and the multimode parameter $M=0.73$, as determined in section \ref{sec:setup}. The fidelity peaks for the expected parameter values of $\eta_{\text{H}}=0.45\pm0.04$ and $\eta_{\text{APD}}=0.45$. Figure \ref{fig:imagesc}(b) shows the fidelity as a function of the ancilla state heralding efficiency $\eta_{\text{H}}$ and the beam splitter reflectivity $R$ for fixed $M=0.73$ and $\eta_{\text{APD}}=0.45$. In the experiment, the beam splitter reflectivity was set to $R=0.5$, i.e. the filter was set to null out the $n=1$ component \cite{Sanaka:PRL:06}. The fidelity peaks in the region of $R=0.5$, adding further confirmation that the reconstructed process tensor is consistent with that predicted by the full model, equation \ref{eq:model_tensor}.

\begin{figure}
\centerline{\includegraphics[width=1.00\textwidth]{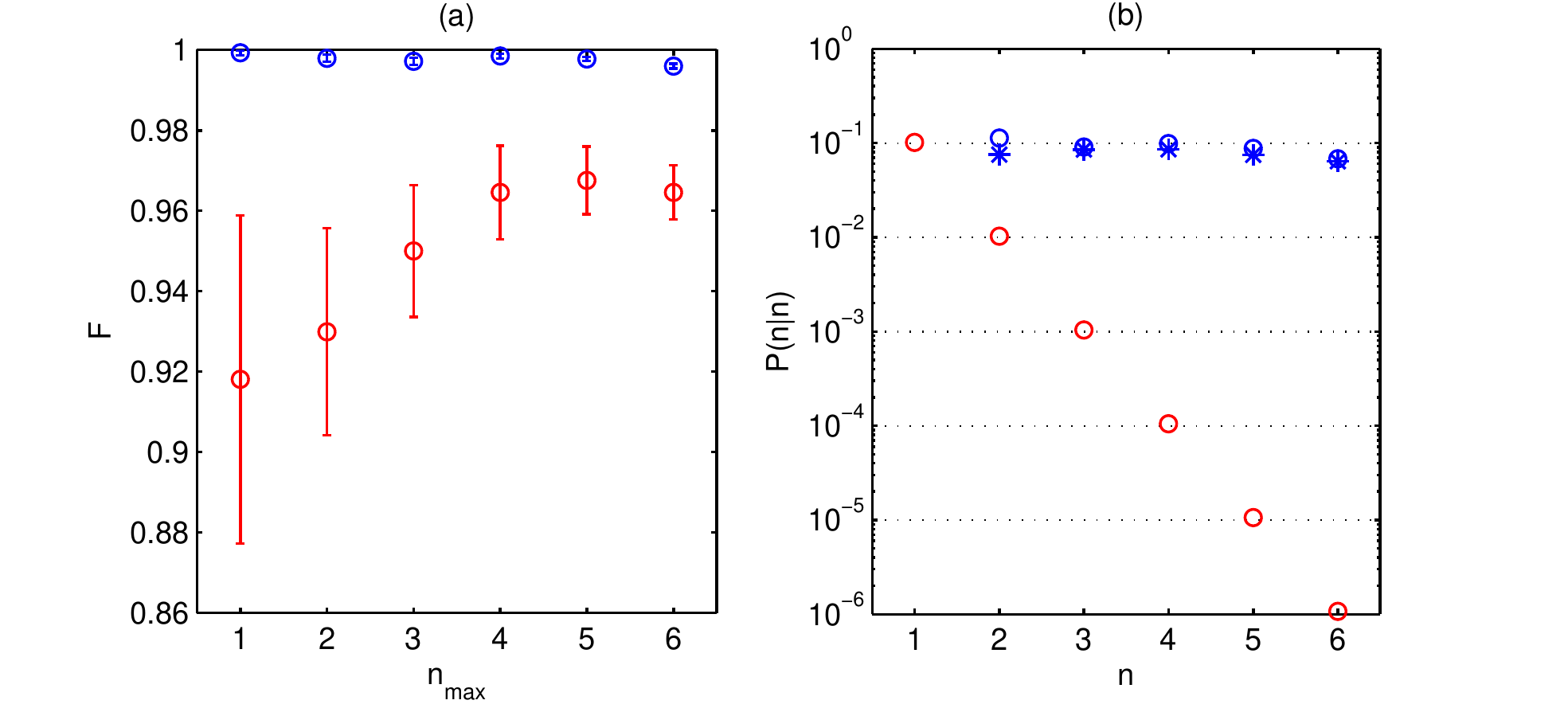}}
\caption{(a) Fidelity between output states due to reconstructed tensor and model tensor as a function of maximum photon number $n_{\text{max}}$ where the model assumed was the full realistic FSF model developed in section \ref{sec:model} (blue) and a simple 50\% attenuation process (red). (b) Probability to obtain $n$ photons at the filter output given $n$ photons were incident to the filter \emph{and} a heralding event is observed, denoted $P(n|n)$. Red circles: based on a linear loss model, blue asterisks: extracted from reconstructed process tensor, blue circles: according to model process tensor. The filter exhibits a nonlinear response since it effectively presents a greater loss to the $n=1$ photon Fock state compared with higher-order Fock states.}
\label{fig:fidelity_attenuation}
\end{figure}
A further test to compare the reconstructed and modeled process tensors is based upon calculating the fidelity between output quantum states predicted by the model and by the reconstructed tensor. To this end, $10^5$ random input quantum states were generated for increasing values of maximum photon number $n_{\text{max}}$ in the state. The random input states were evolved according to the reconstructed and model tensors to give two normalized output states for each random input state, according to
\begin{align}
\hat{\rho}'_{\text{recon}}=\frac{\mathcal{E}_{\text{recon}}(\hat{\rho})}{\text{Tr}\left[\mathcal{E}_{\text{recon}}(\hat{\rho})\right]},\quad {\text{and}}\quad \hat{\rho}'_{\text{model}}=\frac{\mathcal{E}_{\text{model}}(\hat{\rho})}{\text{Tr}\left[\mathcal{E}_{\text{model}}(\hat{\rho})\right]}.
\end{align}
The standard definition for fidelity between two quantum states is employed \cite{Nielsen::10}, similar to equation \ref{eq:6:fidelity_jcoi}, with the process `super-operators,' ${\hat{E}}$, replaced with the state density operators, $\hat{\rho}$. From the set of input states both the mean fidelity and standard deviation could be estimated as a function of $n_{\text{max}}$, blue data in figure \ref{fig:fidelity_attenuation}(a). First, note that the fidelity between the output states is effectively unity when the model process tensor corresponds to the realistic FSF model developed in section \ref{sec:model}, with the parameter values taken as those experimentally determined in section \ref{sec:setup}. As a demonstration that the reconstructed tensor is distinct from the process of $50\%$ attenuation, the same fidelity was calculated where the model tensor was taken to be attenuation by $50\%$, red data in figure \ref{fig:fidelity_attenuation}(a). The results indicate that the full model describing realistic FSF developed in section \ref{sec:model} provides a far better description of the state evolution than a simple 50\% attenuation process. The reason why this result is important can be understood by examining the schematic of FSF in figure \ref{fig:fsf_ideal_model}(b). If the heralding measurement and ancilla state would have no effect on the filter input state then the effect of the operation would simply be to attenuate the input state by a factor $1-R$. Thus by demonstrating that the reconstructed tensor corresponding to the experimental filter is distinct from such an attenuation process the presence of a non-trivial state transformation due to the FSF operation is identified. 

For this experiment the FSF beam splitter reflectivity is $R=0.5$, indicating that the filter should preferentially filter out the $n=1$ photon-number component of the input state $\hat{\rho}$ \cite{Sanaka:PRL:06}. The \emph{un-normalized} photon-number statistics at the filter output given that $n$ photons were incident are given by
\begin{align}
 \tilde{P}(k|n)=\mathcal{E}_{kk}^{nn},
\end{align}
where as usual $\mathcal{E}_{kk}^{nn}$ are the diagonal elements of the process tensor and $\tilde{P}$ denotes the fact that the distribution is not normalized. Summing over $k$ gives the success probability of the FSF operation given that $n$ photons were incident. For the following analysis, the photon-number statistics at the filter output are of interest only when the operation is successful, i.e. a heralding event is observed. Thus the distribution $\tilde{P}(k|n)$ can be normalized according to
\begin{align}
{P}(k|n)=\frac{\tilde{P}(k|n)}{\sum_k \tilde{P}(k|n)},
\end{align}
where ${P}(k|n)$ is the \emph{normalized} photon-number distribution at the filter output given that $n$ photons were incident to the filter \emph{and} a heralding event was observed.

From the reconstructed tensor, $P(1|1)=0.101$, and $P(0|1)=0.899$ -- in other words, if a single-photon is sent into the filter and a heralding event is observed, approximately 9 times out of 10 the filter output state is the vacuum state -- thus the filter exhibits approximately 90\% loss for a single-photon state. If the filter were a passive optical element exhibiting linear loss then one would expect
\begin{align}
P(n|n)=P(1|1)^n, \label{eq:prob_mm}
\end{align}
since for linear loss if a single-photon has a probability $P$ of transmission then a two-photon state has a probability $P^2$ of being preserved and so forth for the higher-order Fock states. This linear loss behaviour is plotted in figure \ref{fig:fidelity_attenuation}(b) (red circles), based on the value $P(1|1)=0.101$ extracted from the reconstructed tensor, and applying equation \ref{eq:prob_mm} to predict the behaviour of the filter if it were considered as linear loss. This is compared against the \emph{actual} values $P(n|n),\,\,n>1$, extracted from the reconstructed process tensor -- plotted as the blue asterisks in figure \ref{fig:fidelity_attenuation}(b). 

Remarkably, the reconstructed process tensor predicts an approximately flat probability of `survival' $P(n|n)$ for the higher-order input Fock states, whereas the values based on a linear loss model decrease factorially. Thus the filter effectively exhibits significantly less loss for Fock states of order $n>1$ than for the $n=1$ Fock state -- consistent with the beam splitter reflectivity $R=0.5$. This demonstrates that the filter is preferentially removing the $n=1$ photon-number component from the input state \cite{Sanaka:PRL:06,Resch:PRL:07} -- i.e. the filter exhibits a nonlinear response to the input state. $P(n|n)$ was also calculated from the model process tensor of equation \ref{eq:model_tensor} using the parameter values obtained in section \ref{sec:setup} -- plotted as the blue circles in figure \ref{fig:fidelity_attenuation}(b). The predicted values for $P(n|n)$ are in excellent agreement with those extracted from the reconstructed tensor and again are completely distinct from the behaviour of linear loss. 

\section{Conclusion}
In this article the photon-level quantum operation known as Fock state filtration \cite{Sanaka:PRL:06} was investigated when implemented with realistic components. A model of the filter operation was developed incorporating experimental imperfections and put to the test by performing quantum process tomography of an experimental implementation of the Fock state filter. Although the practical filter operation was found to be non-ideal, we have identified key challenges to be met for improved performance. The full process tensor reconstruction goes beyond previous demonstrations based on Hong-Ou-Mandel interference \cite{Sanaka:PRL:06,Resch:PRL:07}, and clearly shows the coherence-preserving property and nonlinear behavior of this operation, indicating the potential for this class of operations for quantum state engineering. Three key experimental challenges  were identified -- with the main issue being the unambiguous determination of photon-number in the FSF heralding mode. With the emergence of high-efficiency photon-number-resolving detectors \cite{Hadfield:NatPhoton:09,Calkins:OpEx:13} and improved sources of photonic Fock states \cite{Spring:OpEx:13} it should be possible to achieve near-ideal Fock state filtration and hence also photon-catalysis \cite{Bartley:PRA:12} -- enabling the tailored generation of arbitrary quantum optical states in the continuous-variable domain. The results presented here show that coherent-state quantum process tomography \cite{Rahimi-Keshari:NJP:11,Anis:NJP:12} should serve as a reliable diagnosis tool for such photon-level conditional quantum operations.

\ack
This work was supported by the University of Oxford John Fell Fund and EPSRC grant no. EP/E036066/1. BJS was partially supported by the Oxford Martin School programme on Bio-Inspired Quantum Technologies and EPSRC grant EP/K034480/1. M.K. was supported by a Marie Curie Intra-European Fellowship no. 301032 within the European Community 7th Framework Programme.

\section*{References}
\bibliographystyle{unsrt}
\bibliography{refs}
\end{document}